\documentclass[a4paper,twocolumn,english,prl,superscriptaddress,longbibliography,fixfloat,notitlepage]{revtex4-2}
\usepackage[T1]{fontenc}
\usepackage[latin9]{inputenc}
\usepackage{dsfont}
\usepackage{amsmath}
\usepackage{amssymb}
\usepackage{graphicx}

\makeatletter

\pdfpageheight\paperheight
\pdfpagewidth\paperwidth

\usepackage{xcolor}
\definecolor{darkblue}{rgb}{0.1,0.2,0.6} 
\definecolor{lightblue}{rgb}{0.1,0.1,1.0}
\definecolor{darkred}{rgb}{0.8,0.1,0.2}
\usepackage[colorlinks,citecolor=lightblue,linkcolor=darkblue,urlcolor=lightblue] {hyperref}
\renewcommand{\BibitemShut}[1]{}

\makeatother

\usepackage{babel}
\begin{document}
\global\long\def\E{\mathrm{e}}%
\global\long\def\D{\mathrm{d}}%
\global\long\def\I{\mathrm{i}}%
\global\long\def\mat#1{\mathsf{#1}}%
\global\long\def\vec#1{\mathsf{#1}}%
\global\long\def\cf{\textit{cf.}}%
\global\long\def\ie{\textit{i.e.}}%
\global\long\def\eg{\textit{e.g.}}%
\global\long\def\vs{\textit{vs.}}%
 
\global\long\def\ket#1{\left|#1\right\rangle }%

\global\long\def\etal{\textit{et al.}}%
\global\long\def\tr{\text{Tr}\,}%
 
\global\long\def\im{\text{Im}\,}%
 
\global\long\def\re{\text{Re}\,}%
 
\global\long\def\bra#1{\left\langle #1\right|}%
 
\global\long\def\braket#1#2{\left.\left\langle #1\right|#2\right\rangle }%
 
\global\long\def\obracket#1#2#3{\left\langle #1\right|#2\left|#3\right\rangle }%
 
\global\long\def\proj#1#2{\left.\left.\left|#1\right\rangle \right\langle #2\right|}%
\global\long\def\mds#1{\mathds{#1}}%

\title{Logarithmic, noise-induced dynamics in the Anderson insulator}
\author{Talía L. M. Lezama}
\address{Department of Physics, Ben-Gurion University of the Negev, Beer-Sheva
84105, Israel}
\author{Yevgeny Bar Lev}
\address{Department of Physics, Ben-Gurion University of the Negev, Beer-Sheva
84105, Israel}
\begin{abstract}
We study the dynamical behavior of the Anderson insulator in the presence
of a local noise. We show that the noise induces logarithmically slow
energy and entanglement growth, until the system reaches an infinite-temperature
state, where both quantities saturate to extensive values. The saturation
value of the entanglement entropy approaches the average entanglement
entropy over all possible product states. At infinite temperature,
we find that a density excitation spreads logarithmically with time,
without any signs of asymptotic diffusive behavior. In addition, we
provide a theoretical picture which qualitatively reproduces the phenomenology
of particle transport.
\end{abstract}
\maketitle

\section{Introduction}

Anderson localization is an ubiquitous wave phenomenon that arises
due to destructive interference in the presence of quenched disorder
\citep{Anderson:1958}. Since its discovery, it has been instrumental
for the understanding of a much richer class of physical phenomena
\citep{Lee:rev,Kramer:rev,Fleishman:1980,Basko:2006,Gornyi:2005}.
One of the most important manifestations of Anderson localization
is the suppression of transport which follows from the exponential
localization of all the single-particle wavefunctions in one and two
dimensions \citep{Anderson:1958,Abrahams:1979}, such that infinitesimally
small disorder leads to zero DC-conductivity at any temperature. While
in higher dimensions, and at zero temperature, a metal-insulator transition
takes place \citep{Abrahams:1979}. The phenomenon of Anderson localization
has been extensively studied \citep{Mirlin:2000,Evers:rev}, and experimentally
demonstrated in many systems \citep{Condat:1987,Wiersma:1997,Billy:2008,Roati:2008,Sanchez:2010,Kondov:2011,Jendrzejewski:2012,Segev:2013}.
Even if most of the experimental setups are highly controlled, it
is impossible to completely isolate the system from all dissipation
effects that arise due to the coupling to the environment, for example,
coupling to phonons is present in any condensed matter system, and
is known to induce a finite DC conductivity \citep{Mott:1969}. It
is therefore of importance to theoretically account for such dissipative
processes.

The stability of Anderson localization to different classes of perturbations
has been assessed in several theoretical studies \citep{Fleishman:1980,Altshuler:1997,Bourgain:2002,Basko:2006,Gornyi:2005,Pikovsky:2008,Amir:2009,Amir:2010,Fishman:2012,Huse:2015,Gopalakrishnan:2017,Ducatez:2017,Lorenzo:2018}.
Anderson localization is known to be stable under spatially local
but quasiperiodic in time perturbations in any dimension \citep{Bourgain:2002},
and to survive global periodic driving in one dimension \citep{Agarwal_Anderson:2017,Ducatez:2017}.
In contrast, it is unstable to global noise, which is known to induce
delocalization \citep{Gefen:1984,Logan:1987,Evensky:1990,Evensky:1993,Amir:2009,Amir:2010,Gopalakrishnan:2017,Lorenzo:2018,Taylor:2021},
and a transient subdiffusive transport, which eventually crosses over
to regular diffusion \citep{Gopalakrishnan:2017,Taylor:2021}. Global
noise was also shown to lead to prethermal energy plateaus at intermediate
time scales, followed by exponential relaxation at longer time scales \citep{Lorenzo:2018}.

In this work, we study how the dynamics in the one-dimensional Anderson
insulator is affected by the presence of a \emph{local} white-noise,
which can be thought as a coupling to a local Markovian bath. We find
that the noise leads to a logarithmically slow heating of the system
up to an infinite-temperature state which is further reflected in
slow transport properties of the system.

Our article is organized as follows. In Sec.~\ref{sec:model}, we
introduce the model and the methods used to characterize the noise-induced
dynamics. In Sec.~\ref{sec:dynamics}, we first assess the heating
dynamics in terms of the energy and the entanglement entropy. We then
analyze the particle transport in the system using both numerical
simulations and a semi-analytical approach. In Sec.~\ref{sec:conclusions},
we summarize and discuss our main results.

\section{Model and methods}

\label{sec:model}

We consider the one-dimensional Anderson model, 
\begin{equation}
\hat{H}_{A}=-J\sum_{i=1}^{L-1}\left(\hat{c}_{i}^{\dagger}\hat{c}_{i+1}+\mathrm{h.c.}\right)+\sum_{i=1}^{L}w_{i}\hat{n}_{i},\label{anderson}
\end{equation}
where $\hat{c}_{i}^{\dagger}$ ($\hat{c}_{i}$) creates (annihilates)
a spinless electron on site $i$, ``h.c.'' stands for a Hermitian
conjugate, $\hat{n}_{i}=\hat{c}_{i}^{\dagger}\hat{c}_{i}$ is the
density, and $J$ denotes the hopping constant. The on-site disorder
potential, $w_{i}$, are independent random variables uniformly distributed
in the interval $w_{i}\in\left[-W,W\right]$, with $W$ the disorder
strength. This model is known to be localized for any $W>0$ \citep{Anderson:1958}.
We perturb the Anderson model, by the addition of a local white-noise,
\begin{equation}
\hat{H}=\hat{H}_{A}+\zeta\left(t\right)\hat{n}_{L/2},\label{hamiltonian}
\end{equation}
 such that $\zeta\left(t\right)$ has zero mean $\overline{\zeta(t)}=0$
and a vanishing correlation length 
\begin{equation}
\overline{\zeta\left(t\right)\zeta\left(t'\right)}=\gamma\delta\left(t-t'\right),\label{deltap}
\end{equation}
where the overbar denotes the average over stochastic realizations
of the noise and $\gamma$ is the noise strength. The noise term represents
a local Markovian heat bath coupled to the system, and as such, the
dynamics of the density matrix of the system, $\hat{\rho}\left(t\right)$,
is given by the following Lindblad equation~\citep{Lindblad:1976}
\begin{align}
\partial_{t}\hat{\rho}\left(t\right) & =-i\left[\hat{H}_{A},\hat{\rho}\left(t\right)\right]+\label{lindblad}\\
 & +\gamma\Big(\hat{n}_{L/2}\hat{\rho}\left(t\right)\hat{n}_{L/2}-\frac{1}{2}\left\{ \hat{n}_{L/2},\hat{\rho}\left(t\right)\right\} \Big),\nonumber 
\end{align}
where $\left\{ \cdot,\cdot\right\} $ is the anti-commutator. The
Lindblad equation describes a trace-preserving non-unitary evolution,
where the first term corresponds to the unitary evolution, and the
second term corresponds to the dissipative coupling between the system
and the local heat bath. It is easy to check by substitution, that
the steady state of (\ref{lindblad}) is an infinite-temperature state
with density matrix $\hat{\rho}_{\infty}\propto\mathbb{\mathbb{\mathbb{\mathds{\mathds{1}}}}}$,
which means that for any initial state the system will approach $\hat{\rho}_{\infty}$.
Please note, that the approach to infinite temperature by itself does
\emph{not} imply delocalization of the system, since as stated above,
in one and two dimensions, and without the coupling to the noise,
localization persists also at infinite temperature \citep{Anderson:1958}.
Here, we focus on the questions of \emph{how} the infinite-temperature
state is approached and what is the dynamics of the system at this
state, in the presence of a local noise.

While (\ref{lindblad}) can be numerically solved, this is extremely
demanding even for noninteracting particles, since certain couplings
to the heat bath create an effective interaction between the particles,
which requires the use of the full many-body density matrix of dimensions
$\mathcal{N}\times\mathcal{N}$, where $\mathcal{N}$ is the Hilbert-space
dimension. Alternative methods, based on a unitary propagation followed
by stochastic measurements, of an ensemble of wavefunctions, were
developed \citep{Plenio:rev,Gardiner:2004,Brun:2000,Wiseman:2001,Salgado:2002}.
These methods, known as quantum-trajectory methods, are more efficient
since the dimension of the wavefunction is $\mathcal{N}$. The solution
of (\ref{lindblad}) is reproduced by an average over quantum trajectories,
which correspond to individual realizations of the measurements. The
procedure of writing (\ref{lindblad}) as a stochastic differential
equation, is known as ``unraveling'', and since there can be many
stochastic differential equations whose averages reproduce (\ref{lindblad}),
the procedure is not unique, and can depend on the physical context
\citep{Brun:2000}.

In this work, we use a unitary unraveling of \eqref{lindblad}, which
was introduced in Refs.~\citep{Wiseman:2001,Salgado:2002} and corresponds
to the following stochastic unitary infinitesimal propagator
\begin{equation}
\hat{U}\left(t+\mathrm{d}t,t\right)=e^{-i\hat{H}\mathrm{d}t-i\eta_{t}\hat{n}_{L/2}\sqrt{\gamma dt}},\label{unraveling}
\end{equation}
where $\eta_{t}$ are independent normally distributed random variables.
The evolution of the density matrix is then obtained by performing
an average over trajectories corresponding to the different realizations
of the noise $\eta_{t}$, namely,
\begin{equation}
\hat{\rho}\left(t+\mathrm{d}t\right)=\overline{\ket{\psi\left(t+\mathrm{dt}\right)}\bra{\psi\left(t+\mathrm{dt}\right)}},
\end{equation}
where the overbar denotes the average over the noise trajectories,
and $\ket{\psi\left(t+\mathrm{d}t\right)}=\hat{U}\left(t+\mathrm{d}t,t\right)\ket{\psi\left(t\right)}$,
with the initial condition $\ket{\psi\left(t=0\right)}$ taken from
an ensemble whose average corresponds to the initial density matrix
$\hat{\rho}(t=0)=\overline{\ket{\psi\left(t=0\right)}\bra{\psi\left(t=0\right)}}$.

For self-adjoint Lindblad operators this unraveling is equivalent
to the quantum-jump approach \citep{Salgado:2002}, but is numerically
superior for an initial quadratic density matrix $\hat{\rho}\left(t=0\right)=e^{-\sum_{i}\alpha_{i}\hat{n}_{i}}$,
since it only requires the propagation of a \emph{single-particle}
density matrix 
\begin{equation}
\rho_{ij}^{s}\left(t\right)\equiv\tr\left(\hat{\rho}\left(t\right)\:\hat{c}_{i}^{\dagger}\hat{c}_{j}\right),\label{eq:single-particle-density-matrix}
\end{equation}
which is polynomial rather than exponential in $L$. This simplification
occurs, since $\hat{U}\left(t+\mathrm{d}t,t\right)$ is quadratic
in $\hat{c}_{i}^{\dagger}$ ($\hat{c}_{i}$) and therefore an initially
quadratic density matrix, stays quadratic for the entire evolution
of the system. The propagation of $\rho_{ij}^{s}\left(t\right)$ is
obtained using the single-particle version of the stochastic unitary
propagator (\ref{unraveling}),
\begin{equation}
U^{s}\left(t+\mathrm{d}t,t\right)=e^{-i\hat{h}_{A}\mathrm{d}t-i\eta_{t}\ket{L/2}\bra{L/2}\sqrt{\gamma dt}},
\end{equation}
where
\begin{equation}
\hat{h}_{A}=-J\sum_{i=1}^{L-1}\left(\ket i\bra{i+1}+\ket{i+1}\bra i\right)+\sum_{i=1}^{L}w_{i}\ket i\bra i
\end{equation}
 is the single-particle Anderson Hamiltonian. The evolved single-particle
density matrix $\rho_{ij}^{s}\left(t+\mathrm{dt}\right)$ is therefore
given by,

\begin{equation}
\rho_{ij}^{s}\left(t+\mathrm{d}t\right)=\overline{U_{ik}^{s}\left(t+\mathrm{d}t,t\right)\rho_{kl}^{s}\left(t\right)\hat{U}_{jl}^{s*}\left(t+\mathrm{d}t,t\right)}.
\end{equation}

For our numerical simulations we use Krylov-space methods and time
steps of $\mathrm{d}t=0.1$, which we verified to be sufficient to
obtain converged results. We fix the tunneling constant to $J=1$,
which determines the units of time, and we set the noise strength
to $\gamma=1$. We have seen that changing the amplitude of the noise,
does not change our results qualitatively. We average our results
over 100 disorder realizations and 10 realizations of the noise for
each disorder realization. The averages over disorder are denoted
by $[\cdot]$ and over the noise by an overbar.

\section{Results}

\label{sec:dynamics}

In this section we characterize how the system approaches an infinite-temperature
state in terms of the energy and a properly defined entanglement entropy.
We then assess the linear response particle transport at infinite
temperature.

\subsection{Energy dissipation}

The energy of the system, 
\begin{equation}
\varepsilon\left(t\right)=\tr\left(\hat{\rho}\left(t\right)\,\hat{H}_{A}\right),\label{energy}
\end{equation}
grows as a result of coupling to the local heat bath. Since $\hat{H}_{A}=\sum_{ij}\left\langle i\left|\hat{h}_{A}\right|j\right\rangle \,\hat{c}_{i}^{\dagger}\hat{c}_{j}$
, we can express the energy growth using the single-particle density
matrix as, 
\begin{equation}
\begin{split}\varepsilon\left(t\right) & =\sum_{i,j}\left\langle i\left|\hat{h}_{A}\right|j\right\rangle \tr\left(\hat{\rho}\left(t\right)\hat{c}_{i}^{\dagger}\hat{c}_{j}\right)\\
 & =\sum_{i,j}\rho_{ij}^{s}\left(t\right)\left\langle i\left|\hat{h}_{A}\right|j\right\rangle ,
\end{split}
\end{equation}
where $\ket i$ and $\ket j$ are single-particle states in the position
basis. Following the discussion of Sec.~\ref{sec:model}, at long
times, the system approaches an infinite-temperature state, $\hat{\rho}_{\infty}\propto\mathds{1}$,
therefore,
\begin{equation}
\varepsilon\left(t\to\infty\right)=\frac{1}{\mathcal{N}}\tr\left(\hat{H}_{A}\right)=\frac{1}{2}\tr\hat{h}_{A}=\frac{1}{2}\sum_{i}^{L}w_{i},
\end{equation}
where $\mathcal{N}$ is the Hilbert space dimension. Since the energy
of the system is bounded, to have a sufficiently wide range of energy
growth, we prepare the system in the ground state of $\hat{H}_{A}$.
In this state, the single-particle density matrix is given by,
\begin{equation}
\rho_{ij}^{s}\left(0\right)=\sum_{\alpha=1}^{N}\phi_{\alpha}^{*}\left(i\right)\phi_{\alpha}\left(j\right),\label{idm}
\end{equation}
where $\ket{\alpha}$ are single-particle eigenstates of $\hat{h}_{A}$
and $N$ is the number of fermions, which we set to be $N=L/2$, namely,
half-filling.

\begin{figure}[tb!]
\centering{}\includegraphics[width=1\columnwidth]{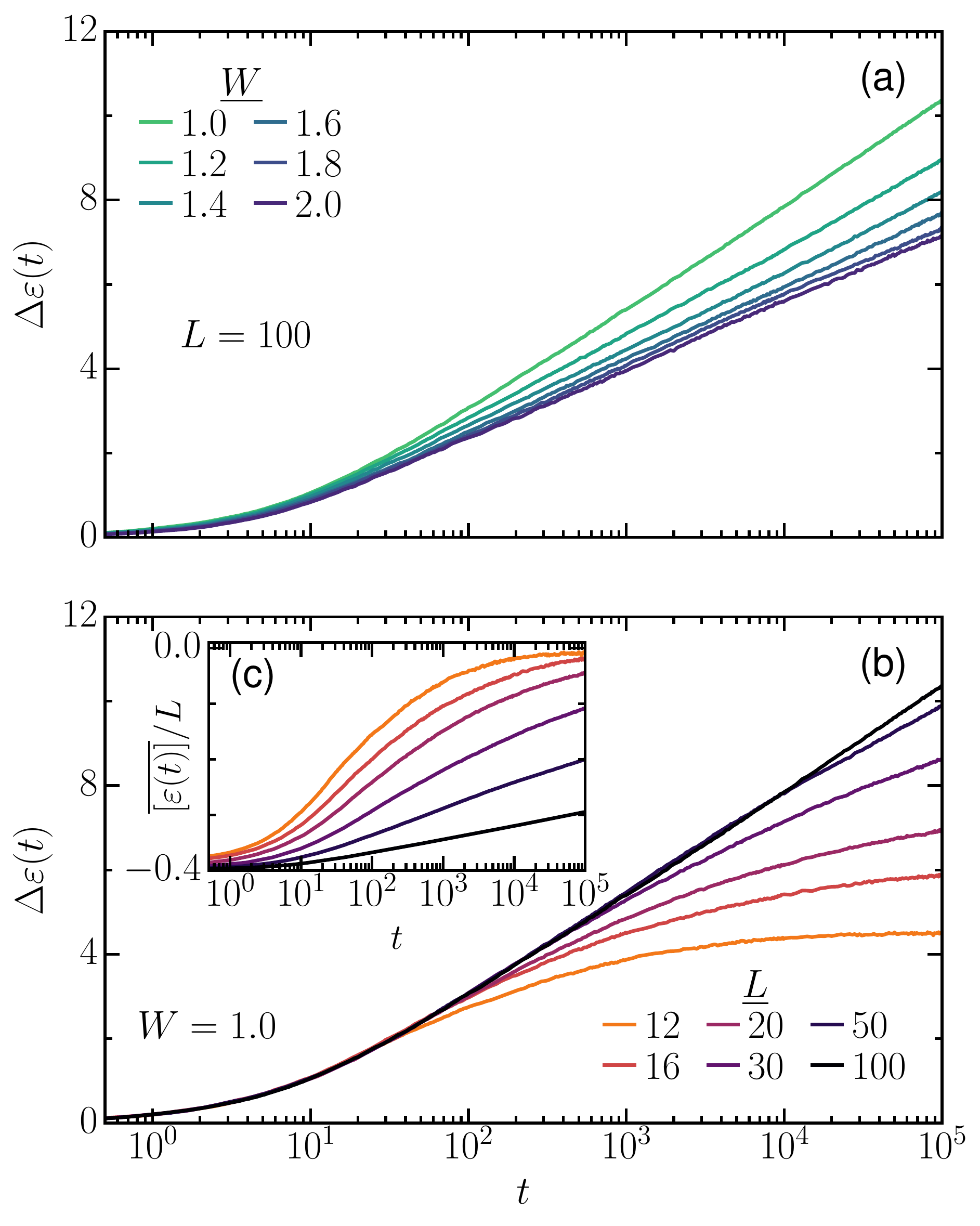} \caption{Averaged energy growth. (a) As a function of disorder strength $W$
for $L=100$, and (b) as a function of system size $L$ for $W=1.0$.}
\label{fig:energy}
\end{figure}

In Fig.~\ref{fig:energy}(a) we show how the averaged (over realizations
of disorder and trajectories) energy absorbed from the coupling to
the local environment, $\Delta\varepsilon\left(t\right)=\left[\overline{\varepsilon\left(t\right)-\varepsilon_{\mathrm{GS}}}\right]$,
grows in time for several disorder strengths $W$ and a fixed system
size $L=100$. We observe that the averaged absorbed energy grows
logarithmically with time, $\Delta\varepsilon\sim\ln t$, over a broad
time window extending into several decades and for all the disorder
strengths we study. Increasing the disorder strength further suppresses
the heating. Interestingly, this logarithmically slow energy growth
resembles the heating in a vicinity of the Floquet-MBL transition
\citep{Rehn:2016}.

In Fig.~\ref{fig:energy}(b), we show how $\Delta\varepsilon\left(t\right)$
depends on system size $L$, for a weak disorder $W=1$. For the smallest
system size $L=12$, the absorbed energy saturates, but as we increase
the system size, the times required to observe saturation significantly
increase. Since the energy of the system is extensive, to compare
the saturation of the energy for different system sizes we calculate
the energy density $\varepsilon(t)/L.$ The results are presented
in the inset of Fig.~\ref{fig:energy}(b) and show the approach to
an energy density corresponding to an infinite-temperature state,
which for $\hat{H}_{A}$ is $\varepsilon_{\infty}/L=\tfrac{1}{2}\left\langle w_{i}\right\rangle =0$.

\subsection{Entanglement entropy growth}

The entanglement entropy is not well defined for mixed states \citep{Eisert_entanglement:rev},
and therefore is not a natural quantity to consider for dissipative
dynamics given by (\ref{lindblad}). However, since (\ref{lindblad})
can be unraveled into a unitary evolution of an ensemble of pure states
(see Sec.~\ref{sec:model}), a well-defined entanglement entropy
can be computed for each of the members of the ensemble \emph{separately},
and then averaged. Since the entanglement entropy is not a linear
function of the wavefunctions, the averaged entanglement entropies
corresponding to different unravelings of (\ref{lindblad}) do \emph{not}
need to coincide \citep{Cao:2019}. Notwithstanding, the averaged
entanglement can obtain meaning, if a certain unraveling can be physically
motivated \citep{Gardiner:2004}. Specifically, the unitary unraveling
of Refs.~\citep{Wiseman:2001,Salgado:2002} that we use here, can
be thought of as a time-dependent, multi-frequency local potential,
with a band-width much larger than any other energy scale.

To calculate the von-Neumann entanglement entropy we partition the
system into two spatially equal parts $A$ and $B$. For noninteracting
systems the entanglement between $A$ and $B$ is given by~\citep{Peschel:2009}
\begin{equation}
\begin{split}S\left(t\right)= & -\sum_{\alpha}\Big[\tilde{n}_{\alpha}^{A}\left(t\right)\log\tilde{n}_{\alpha}^{A}\left(t\right)\\
 & +\left(1-\tilde{n}_{\alpha}^{A}\left(t\right)\right)\log\left(1-\tilde{n}_{\alpha}^{A}\left(t\right)\right)\Big],
\end{split}
\label{entropy}
\end{equation}
where $\tilde{n}_{\alpha}^{A}\left(t\right)$ are the eigenvalues
of the single-particle density matrix $\rho_{ij}^{s}\left(t\right)$
with $i,j\in A$. To have a sizable regime of entanglement growth
we initiate the system at a random product state, namely, $\rho_{ij}^{s}\left(t=0\right)=n_{j}\delta_{ij}$,
with random $n_{j}\in\left\{ 0,1\right\} $.

\begin{figure}[tb!]
\centering{}\includegraphics[width=1\columnwidth]{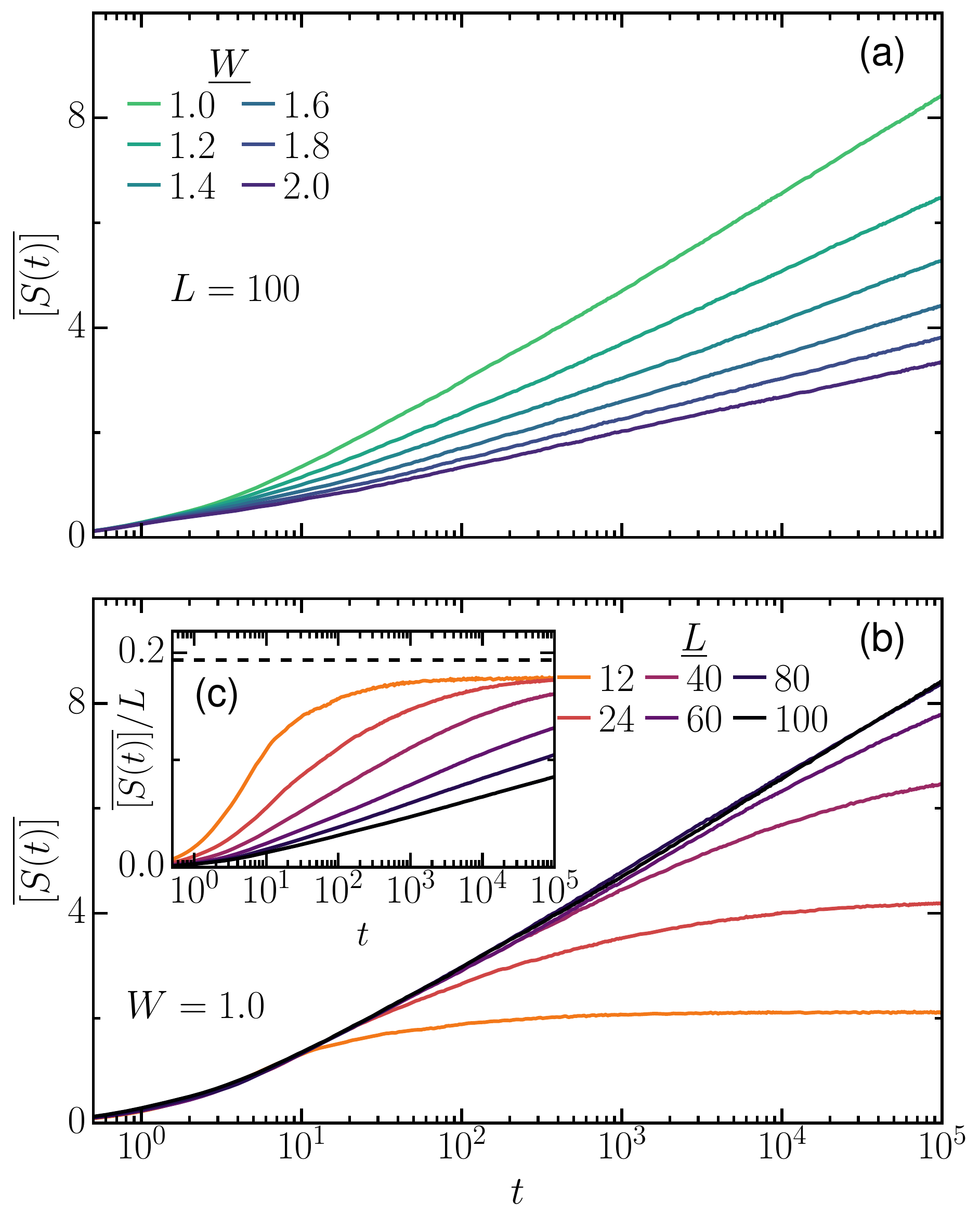}\caption{Averaged entanglement growth. (a) As a function of disorder strength
$W$ for $L=100$, and (b) as a function of system size $L$ for $W=1.0$.
(c) Inset showing the saturation of the averaged entanglement entropy
density. The dashed line in (c) refers to the entanglement entropy
density averaged over all possible product states, $\bar{\mathcal{S}}/L\approx0.193$
(see Eq.~(2) in Ref.~\citep{Lydzba:2020}).}
\label{fig:entropy}
\end{figure}

In Fig.~\ref{fig:entropy}(a), we show the time evolution of the
entanglement entropy $S\left(t\right)$, averaged over disorder and
trajectories realizations, for various disorder strengths $W$, and
a fixed system size $L=100$. Similar to the energy, we find that
the averaged entropy grows logarithmically with time and that the
slope of its growth is suppressed with disorder strength. Since the
entanglement is bounded, the growth saturates at long times to $S_{\infty}\equiv\lim_{t\to\infty}S\left(t\right)$,
as is apparent for the smaller system sizes in Fig.~\ref{fig:entropy}(b).
The entanglement entropy density $S\left(t\right)/L$, for different
system sizes approaches the same constant, $S_{\infty}/L\approx\tfrac{1}{4}\ln2,$
as is shown in the inset of Fig.~\ref{fig:entropy}(b). This value
is considerably smaller than the Page value $S_{\text{Page}}=\tfrac{1}{2}L\ln2-\tfrac{1}{2}$
\citep{Page:1993}, contrary to the case of coupling to noise in \emph{interacting}
systems \citep{Levi:2016}. Since in our case the state of the system
is a product state for all times, though not necessarily in the position
basis, the saturation value better agrees with the entanglement entropy
density averaged over all possible product states, given by $\bar{\mathcal{S}}/L\approx0.193$
(see Eq.~(2) in Ref.~\citep{Lydzba:2020}), and not over all possible
states in the entire Hilbert space, which would correspond to the
Page value. Before concluding this section, it is worthwhile to observe
that the entire behavior of the averaged entanglement entropy in Fig.~\ref{fig:entropy}
is somewhat reminiscent of the entanglement entropy behavior in many-body
localized systems~\citep{Znidaric:2008,Bardarson:2012}, though,
as we will see in what follows, here the system is delocalized by
the noise, which induces a slow particle transport.

\subsection{Particle transport at infinite temperature}

In the previous subsections we have studied the approach of the system
to the infinite-temperature state. Here, we consider particle transport
at infinite temperature. Since in the absence of noise, all single-particle
states are localized, there is no transport, even at infinite temperature.
Therefore, all transport is induced by the local noise. To assess
particle transport we compute the density-density correlation function
\begin{equation}
C_{ij}\left(t\right)=\tr\left(\hat{\rho}_{\infty}\:\hat{n}_{i}\left(t\right)\hat{n}_{j}\right)=\left|\tr\left(\hat{\rho}_{\infty}\:\hat{c}_{i}^{\dagger}\left(t\right)\hat{c}_{j}\right)\right|^{2},
\end{equation}
which corresponds to the spreading of an excitation of the density
at site $j$. The last equality follows from the fact that the particles
are noninteracting, and $\hat{\rho}_{\infty}\propto\mathds{1}$. For
the unitary unraveling of (\ref{lindblad}) that we use here, the
evolution of $\hat{c}_{i}^{\dagger}\left(t\right)$ is given by $\hat{c}_{i}^{\dagger}\left(t\right)=\sum_{k}U_{ik}^{s}\left(t,0\right)\hat{c}_{k}^{\dagger}$,
therefore,
\begin{equation}
C_{ij}\left(t\right)=\left|\sum_{k}U_{ik}^{s}\left(t,0\right)\rho_{kj}^{s}\right|^{2}=\frac{1}{4}\left|U_{ij}^{s}\left(t,0\right)\right|^{2},
\end{equation}
where we used the infinite-temperature form of the single-particle
density matrix, $\rho_{kl}^{s}=\tfrac{1}{2}\delta_{kl}$. To characterize
the nature of transport in the presence of the local noise, we first
evaluate the width of the excitation profile, known as the root-mean-square
(RMS) displacement, 
\begin{equation}
\tilde{R}\left(t\right)=\left(\sum_{i=1}^{L}\left(i-j\right)^{2}\overline{\left[C_{ij}\left(t\right)\right]}\right)^{1/2}.
\end{equation}
For diffusive transport, $\tilde{R}\left(t\right)\sim\sqrt{2Dt}$,
where $D$ is the linear response diffusion constant \citep{Steinigeweg:2009,Steinigeweg:2017},
and for localization the width is bounded, $\tilde{R}\left(t\right)\leq A$.

\begin{figure}[tb!]
\centering{}\includegraphics[width=1\columnwidth]{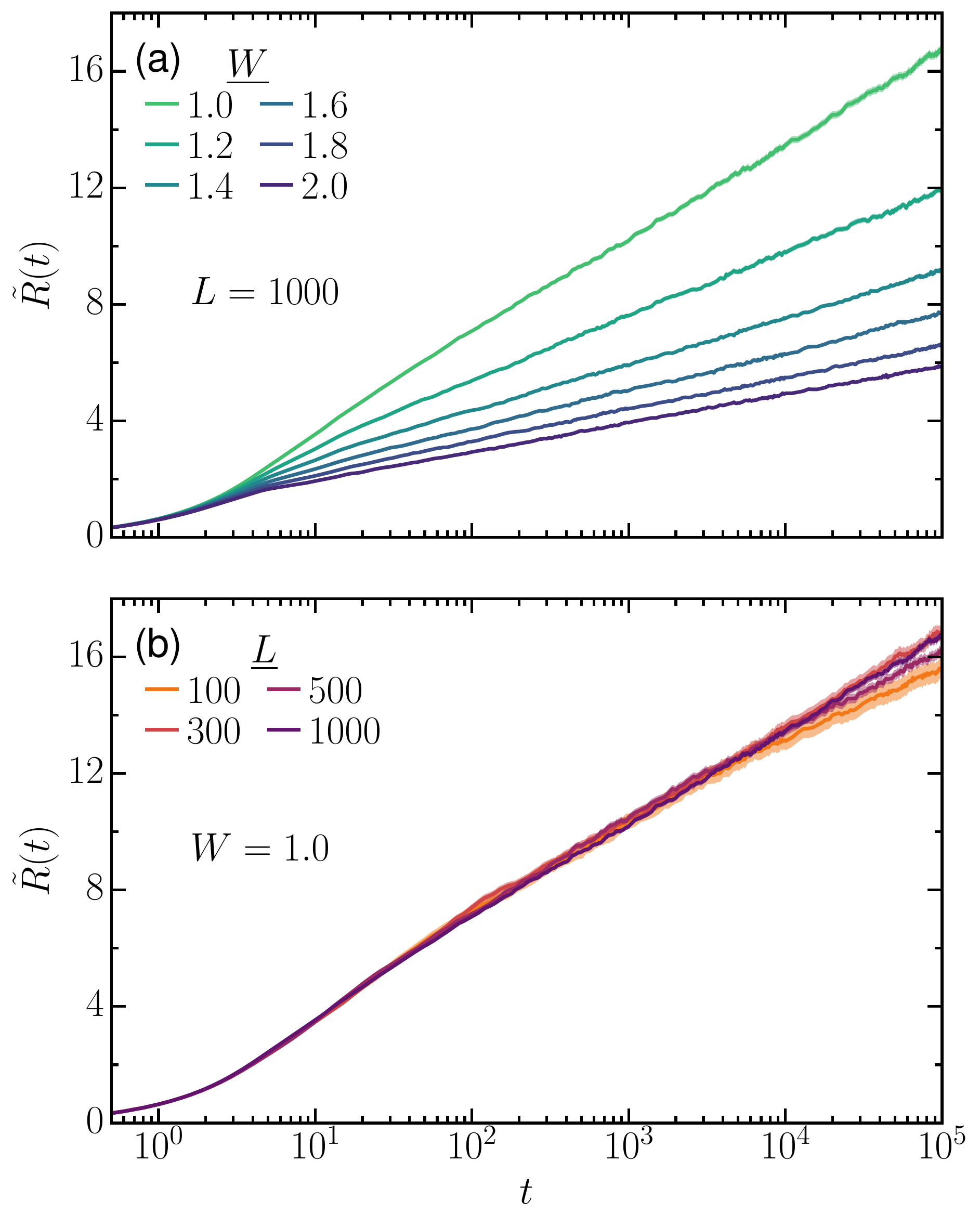} \caption{RMS displacement $\tilde{R}\left(t\right)$ as a function of time.
(a) For various disorder strengths $W$ and $L=1000$, and (b) for
various system sizes $L$ and $W=1.0$.}
\label{fig:msd}
\end{figure}

In Ref.~\citep{Gopalakrishnan:2017}, it was shown that the Anderson
insulator subject to \emph{global} noise with arbitrary correlation
time exhibits transient subdiffusion, before asymptotic diffusion
takes place. In contrast, in the case of local white noise, we find
that the RMS displacement $\tilde{R}\left(t\right)$ grows logarithmically
with time, without any signs of crossover to diffusion (see Fig.~\ref{fig:msd}(a)).
Similarly to the energy and the entanglement entropy, transport is
suppressed with increasing the disorder strength. In Fig.~\eqref{fig:msd}(b)
we show that our results do not suffer from finite-size effects over
a broad time window spanning several decades.

\section{Semi-analytical picture}

\begin{figure}[tb!]
\centering{}\includegraphics[width=1\columnwidth]{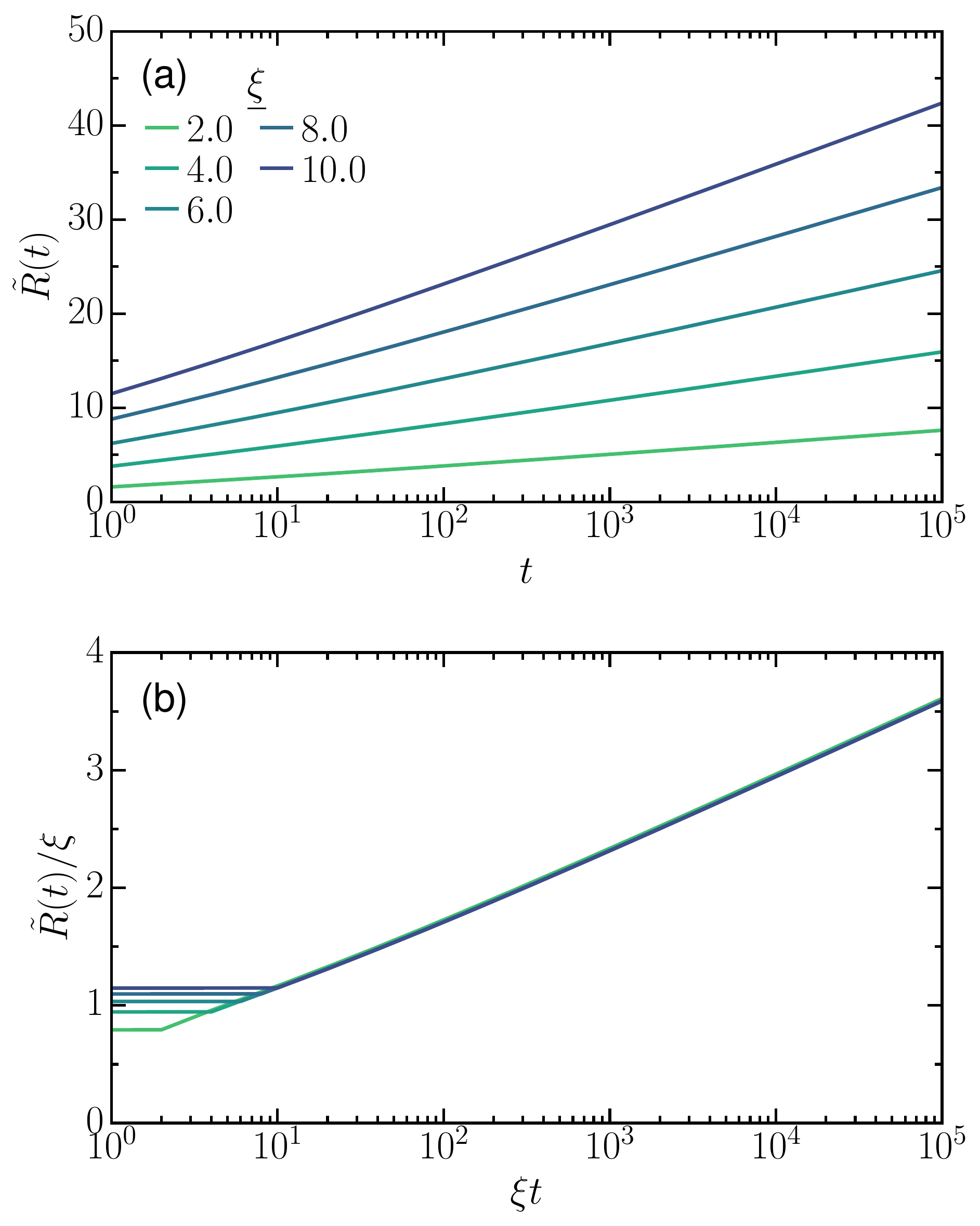} \caption{RMS displacement $\tilde{R}\left(t\right)$ as a function of time.
(a) For various localization lengths $\xi$ and $L=1000$. (b) The
same as (a), but $\tilde{R}\left(t\right)/\xi$ plotted vs $\xi t$.}
\label{fig:msdanalytical}
\end{figure}

In this section we provide a theoretical model of nonequilibrium dynamics
in the Anderson insulator in the presence of a local noise, which
gives a qualitative explanation of the phenomenology we observe. For
this purpose we use a variable-range-hopping-like approach, as originally
introduced by Mott in the context of phonons~\citep{Mott:1969}.
In this approach, the environment induces hopping between the localized
orbitals of the Anderson problem. Moreover, it is assumed that the
noise decoheres the dynamics, such that the process can be described
by a classical master equation \citep{Amir:2009,Amir:2010,Fischer:2016},
\begin{equation}
\partial_{t}p_{\alpha}=\sum_{\beta}\left(\Gamma_{\alpha\beta}p_{\beta}-\Gamma_{\beta\alpha}p_{\alpha}\right),\label{masterequation}
\end{equation}
where $p_{\alpha}$ is the probability to find a particle at an Anderson
eigenstate $\braket i{\alpha}=\phi_{\alpha}\left(i\right)$, and 
\begin{equation}
\begin{gathered}\Gamma_{\alpha\beta}=\Gamma_{\beta\alpha}=\gamma^{2}\left|\braket{\beta}{L/2}\braket{L/2}{\alpha}\right|^{2}\\
=\gamma^{2}\left|\phi_{\beta}^{*}\left(\frac{L}{2}\right)\phi_{\alpha}\left(\frac{L}{2}\right)\right|^{2},
\end{gathered}
\end{equation}
are the transition rates between Anderson eigenstates $\ket{\alpha}$
and $\ket{\beta}$, where to obtain the rates we used the noise coupling
$\gamma\ket{L/2}\bra{L/2}$, and the fact that the noise is white.
Since the Anderson eigenstates are localized, in a one-dimensional
lattice the indices $\alpha$ can be ordered almost in one-to-one
correspondence with the site indices $i$, thus we can write,
\begin{equation}
\Gamma_{\alpha\beta}=\gamma^{2}e^{-\left|\alpha-L/2\right|/\xi}e^{-\left|\beta-L/2\right||/\xi},
\end{equation}
where $\xi$ is the localization length. We see that the transition
rates between a pair of states are exponentially suppressed with the
distance from the local noise, which explains the exponentially long-time
scales we observe. To see this more precisely, we numerically solve
\eqref{masterequation} for a particle initially located at the center
of the lattice. In this case the RMS displacement is given by $\tilde{R}(t)=\sqrt{\sum_{\alpha}\left(\alpha-\frac{L}{2}\right)^{2}p_{\alpha}\left(t\right)}$,
and is plotted in Fig.~\ref{fig:msdanalytical}(a) for various localization
lengths, $\xi$. Plotting $\tilde{R}\left(t\right)/\xi$ with respect
to $\xi t$ results in a perfect collapse of the data, as shown in
Fig.~\ref{fig:msdanalytical}(b), indicating that the RMS displacement
scales as $\tilde{R}\left(t\right)\sim\xi\ln\left(\xi t\right)$,
which is in excellent agreement with the quantum simulation in the
previous section, though we could not produce a similar collapse for
the original problem (\ref{hamiltonian}) with $\xi$ computed numerically.
This might indicate that more than one scaling parameter might be
required.

\section{Discussion}

\label{sec:conclusions}

We have studied the dynamical behavior of the Anderson insulator in
the presence of a local noisy potential. While the dynamics is dissipative,
it can be efficiently studied using an ensemble of pure states, which
evolve under \emph{unitary} evolution \citep{Wiseman:2001,Salgado:2002}.
Physically, this corresponds to dynamics in the presence of a local,
time-dependent potential with a very wide bandwidth of frequencies,
which allows us to consider, in addition to the energy absorption,
the growth of the entanglement entropy. We find that both quantities
grow logarithmically in time and saturate after times that diverge
exponentially with system size. While the local noise leads to an
infinite-temperature state at long times, the entanglement entropy
saturates to an extensive value which is smaller than the Page value \citep{Page:1993}, but that is in good agreement with the average entanglement entropy over all product states \citep{Lydzba:2020}. Interestingly, the entanglement
entropy growth we observe is similar to that of many-body localized
systems \citep{Znidaric:2008,Bardarson:2012}, but contrary to MBL
systems the local noise induces slow logarithmic particle transport
at infinite temperature. This scenario is also different from the
case of global coupling to noise, where subdiffusive transport is
only a transient and asymptotically the system is diffusive \citep{Gopalakrishnan:2017}.

We show that the slow dynamical behavior of the Anderson insulator
in the presence of a local noise, can be qualitatively understood
using a classical master equation, which describes noise-mediated
hopping of a particle between localized single-particle states, similar
to the variable-range-hopping mechanism~\citep{Mott:1969,Amir:2009,Amir:2010,Fischer:2016}.
In particular, we show that the RMS displacement of the particle grows
as, $\tilde{R}\left(t\right)\sim\xi\ln\xi t$, indicating a vanishing
diffusion coefficient. Based on this analysis, it is easy to see that
our results should hold for any noise which operates in a bounded
spatial region, however, the system will become delocalized if the
noise operates on a finite fraction of the lattice, $p$. In this
case, the average distance between the noisy sites would be $\ell=1/p$,
and the system would delocalize in a time scale, $t\sim\xi^{-1}\exp\left[\ell/\xi\right]$,
exhibiting diffusive transport \citep{Gopalakrishnan:2017,Taylor:2021}.
Our results provide an upper bound for the delocalization rate of
Anderson and MBL systems in the presence of local ergodic grains,
discussed in Refs.~\citep{Luitz_bath:2017,Khemani_critical:2017},
since unlike the grains the local noise does not ``cool down''.
\begin{acknowledgments}
We thank Lev Vidmar for bringing to our attention Ref.~\citep{Lydzba:2020},
where the average entanglement entropy over all possible product
states is obtained analytically. This research was supported by a
grant from the United States-Israel Binational Foundation (BSF, Grant
No. 2019644), Jerusalem, Israel, and the United States National Science
Foundation (NSF, Grant No. DMR-1936006), and by the Israel Science
Foundation (grants No. 527/19 and 218/19). TLML acknowledges funding
from the Kreitman fellowship.
\end{acknowledgments}

\bibliography{references}

\end{document}